\documentclass[11pt]{iopart}

\usepackage{amsbsy,amssymb}
\usepackage{mathrsfs}

\newcommand{\eps}{\epsilon}
\newcommand{\al}{\alpha}
\newcommand{\be}{\beta}
\newcommand{\ga}{\gamma}
\newcommand{\de}{\delta}
\newcommand{\m}{\mu}
\newcommand{\n}{\nu}

\newcommand{\si}{\sigma}
\newcommand{\en}{n}
\newcommand{\cj}{\mathscr{J}}
\newcommand{\emrho}{\rho_e}
\newcommand{\nn}{\nonumber}
\newcommand{\mop}{\mathcal{M}_\oplus}
\newcommand{\moc}{\mathcal{M}_\otimes}
\newcommand{\Si}{\Sigma}
\newcommand{\ch}{\mc{H}}
\newcommand{\ta}{\tau}

\newcommand{\mc}{\mathcal}
\newcommand{\ce}{\mc{E}}
\newcommand{\cs}{N}

\newcommand{\ca}{\mc{A}}
\newcommand{\cq}{\mc{Q}}

\newcommand{\na}{\nabla}

\newcommand{\emce}{\mathscr{E}}
\newcommand{\emcb}{\mathscr{B}}

\begin{document}

\title[1+1+2 Electromagnetic perturbations on general LRS space-times]{1+1+2 Electromagnetic perturbations on general LRS space-times: Regge-Wheeler and Bardeen-Press equations}

\author{R. B. Burston$^{1, 2}$ and A. W. C. Lun$^1$}

\address{School of Mathematical Sciences, Monash University, Australia 3800$^1$}
\address{Max Planck Institute for Solar System Research,
37191 Katlenburg-Lindau, Germany$^2$}
\eads{\mailto{burston@mps.mpg.de$^1$}}

\begin{abstract}
We use the, covariant and gauge-invariant, 1+1+2 formalism developed by Clarkson and Barrett \cite{Clarkson2003}, and develop new techniques, to decouple electromagnetic (EM) perturbations on arbitrary {\it locally rotationally symmetric} (LRS) space-times.  Ultimately, we derive 3 decoupled complex equations governing 3 complex scalars. One of these is a new Regge-Wheeler (RW) equation generalized for LRS space-times, whereas the remaining two are new generalizations of the Bardeen-Press (BP) equations. This is achieved by first using linear algebra techniques to rewrite the first-order Maxwell equations in a new complex 1+1+2 form which is conducive to decoupling. This new complex system immediately yields the generalized RW equation, and furthermore, we also derive a decoupled equation governing a newly defined  complex EM 2-vector. Subsequently, a further decomposition of the 1+1+2 formalism into a 1+1+1+1 formalism is developed, allowing us to decompose the complex EM 2-vector, and its governing equations, into spin-weighted scalars, giving rise to the generalized BP equations.
\end{abstract}

\maketitle

\section{Introduction}
The 1+3 formalism is well established (for example, see \cite{Bel1958,Ellis1967,Ehlers1993}), whereby a four-velocity, $u^\m$, is defined such that it is both time-like and normalized, $u^\al u_\al := -1$. Any tensor may then be irreducibly decomposed into several parts; a temporal part in the direction of the time-like four-velocity, a spatial part (referred to as a 3-tensor) which is projected onto the instantaneous rest-space (or 3-sheet) orthogonal to $u^\m$, and parts which comprise combinations of both projections and contractions in the time-like direction (also 3-tensors). Recently, Clarkson and Barrett  developed a further decomposition of the 1+3 formalism into a 1+1+2 formalism for an analysis of gravitational perturbations on a covariant Schwarzschild space-time \cite{Clarkson2003}. They introduced a radial vector field, $n^\m$, defined such that it is space-like and normalized, $n^\al n_\al:=1$, and it is orthogonal to the time-like vector field, $u^\al n_\al =0$. In this way, every 3-tensor may be further irreducibly decomposed into a radial part in the direction of the radial vector, a part which is projected onto a 2-sheet (called a 2-tensor) orthogonal to both $u^\m$ and $n^\m$, and parts which comprise combinations.

The 1+1+2 formalism is an ideal setting for describing electromagnetic (EM) perturbations on {\it locally rotationally symmetric} (LRS) space-times \cite{Ellis1967,Betschart2004,Elst1996,Stewart1968}. EM perturbations via the 1+1+2 formalism were first considered in \cite{Betschart2004} for  {\it LRS class II} space-times, which comprise a sub-set of LRS space-times and this is discussed further in Section \ref{LRSII}. In Section \ref{newstuff},  the necessary background constraint, evolution, propagation and transportation equations are presented for arbitrary LRS space-times, including a full description of energy-momentum sources.

In Section \ref{emPerturbations},  EM perturbations on arbitrary LRS space-times are considered and the corresponding first-order Maxwell equations in 1+1+2 form are reproduced from \cite{Clarkson2004}.  Subsequently, we use eigenvalue/eigenvector analysis to reveal that a very natural way to decouple the 1+1+2 Maxwell equations is to construct new complex dependent variables. Consequently, we display Maxwell's equations in a new 1+1+2 complex form which is conducive to decoupling. Then in Section \ref{RWLRS} we derive a {\it Regge-Wheeler} (RW) equation \cite{Regge1957}, generalized towards arbitrary LRS space-times for a complex scalar.  In Section \ref{BardeenPressLRS}, we derive a decoupled equation governing a complex 2-tensor.  Subsequently, in order to decouple the individual components, we again using linear algebra techniques to best exploit the inherent structure of the equations. Consequently, a further decomposition of the 1+1+2 formalism into a new 1+1+1+1 formalism is developed and the complex 2-tensor $\Phi_\m$ is irreducibly decomposed into two spin-weighted scalars \cite{Penrose1987}. Ultimately, we arrive at a generalization of the {\it Bardeen-Press} (BP) equations \cite{Penrose1987,Bardeen1972, Chandra1983} for LRS space-times. 

Finally, unless otherwise stated, we adhere to the notation employed in \cite{Clarkson2003,Betschart2004}.

\section{The Background LRS space-time}\label{newstuff}

There is a set of 1+1+2 scalar quantities describing arbitrary LRS space-times (also noted in \cite{Betschart2004}) which are given by, 
\begin{eqnarray}
\mbox{LRS:}\,\, \{\ca, \,\theta,\, \phi,\,\Si, \xi,\,\Omega,\,\ce,\ch \,,\,\m,\, p,\,\cq,\,\Pi,\, \Lambda\}. \label{LRSscalars}
\end{eqnarray}
Here, $\ca $ is the radial acceleration of the four-velocity, $\theta$ and $\phi$ are respectively the expansions of the 3-sheets and 2-sheets, $\Si$ is the radial part of the shear of the 3-sheet, $\xi$ is the twisting of the 2-sheet and $\Omega$ is the radial part of the vorticity of the 3-sheet. Also, the radial parts of the gravito-electric and gravito-magnetic tensors are respectively, $\ce$ and $\ch$. The energy-momentum quantities, mass-energy density, pressure, radial heat flux and radial anisotropic stress are denoted respectively $\m$, $p$, $\cq$ and $\Pi$, and finally, $\Lambda$ is the cosmological constant. The covariant 2-derivative associated with the 2-sheets is denoted $\de_\m$ and by operating on any LRS scalar will yield zero.  The 1+1+2 coupled system which governs these scalars arise from the Ricci identities for the vector fields, $u^\m$ and $n^\m$ and the Bianchi Identities. They were presented in \cite{Clarkson2003} for the vacuum Schwarzschild case and in \cite{Betschart2004} for LRS class II space-times. Here, we generalize them further for arbitrary LRS space-times and the system becomes significantly more complicated. Furthermore, an independent study of these equations is carried out in a  very recent paper \cite{Clarkson2007}. Compared to LRS class II space-times, there is a combined total of 7 additional evolution, propagation and constraint equations arising which govern the additional scalar quantities. First, the Ricci and Bianchi identities are defined according to
\begin{eqnarray}
Q_{\m\n\si} &:=2&\,\na_{[\m} \na_{\n]} u_\si - R_{\m\n\si\ta} u^\ta =0,\label{ricciforlittlenm}\\
R_{\m\n\si} &:= 2&\, \na_{[\m} \na_{\n]} \en_\si - R_{\m\n\si\ta} \en^\ta=0\label{riccifornm},\\
B_{\n\si\ta} &:= &\,\na^\m C_{\m\n\si\ta} -  [\na_{[\si} T_{\ta]\n} +\frac 13\, g _{\n[\si} \na_{\ta]}T]=0\label{bianchids},
\end{eqnarray} 
where $C_{\m\n\si\tau}$ is the Weyl tensor, $R_{\m\n\si\ta}$ is the Riemann tensor, $T_{\m\n}$ the energy-momentum tensor and $\na_\m$ the four-dimensional covariant derivative. The governing equations are then categorized into specific groups according to
\subsubsection*{Constraint}\footnote{They are derived as follows. \eref{heqs} from $\eps^{\m\n} u^\si R_{\m\n\si}=0.$} 
  \begin{eqnarray}
3\,\xi\,\Si -(2\,\ca-\phi) \Omega -\ch=0 \label{heqs}
\end{eqnarray} 
\subsubsection*{Propagation}\footnote{Derived as follows.  \eref{rb1} from  $\en^\m \cs^{\n\si} R_{\m\n\si} =0$; \eref{lensimthe} from $\en^\m\cs^{\n\si} Q_{\m\n\si} =0$;  \eref{lence} from $u^\al u^\be \en^\ga B_{\al\be\ga}=0$; \eref{lench} from $\eps^{\be\ga} u^\al B_{\al\be\ga} =0$; \eref{lenxi} from $\en^\m \eps^{\n\si}  R_{\m\n\si}=0$; \eref{lenomega} from $\eps^{\m\n\si} Q_{\m\n\si}=0$.} 
\begin{eqnarray}
\hat\phi+\frac 12\,\phi^2+\Bigl(\Si-\frac 23\,\theta\Bigr)\Bigl(\Si+\frac 13\,\theta\Bigr)+\ce-2\,\xi^2=-\frac 23\,(\mu+\Lambda) -\frac 12\,\Pi, \label{rb1}\\
\hat\Si -\frac23\,\hat\theta+\frac 32\, \phi\,\Si+2\,\xi\,\Omega=-\cq,\label{lensimthe}\\
 \hat \ce+\frac 32\,\phi\,\ce-3\,\ch\,\Omega=\frac13\, \hat \m+\frac 12\,\Bigl( \Si-\frac23\,\theta\Bigr)\,\cq-\frac 12\,\hat \Pi -\frac 34\,\phi \,\Pi,\label{lence}\\
\hat \ch +\frac 32\,\phi\,\ch+3\,\ce\,\Omega=-\Bigl(\m+p-\frac 12\,\Pi\Bigr)\,\Omega-\cq\,\xi,\label{lench}\\
\hat \xi +\phi\, \xi -\Bigl(\Si+\frac 13\,\theta\Bigr)\Omega =0\label{lenxi}, \\
\hat \Omega -(\ca-\phi) \,\Omega =0 .\label{lenomega}
\end{eqnarray} 
\subsubsection*{Evolution}\footnote{Derived as follows. \eref{luphi} from $u^\m \,\cs^{\n\si} R_{\m\n\si} =0$;  \eref{lusimthe} from   $u^\m \cs^{\n\si} Q_{\m\n\si}=0$; \eref{luce} from $  u^\al  \en^\be \en^\ga B_{\be\ga\al}=0$;  \eref{luch} from $ \eps^{\al\be} \en^\ga B_{\ga\al\be}=0$; \eref{luxi} from $u^\m \eps^{\n\si}  R_{\m\n\si}=0$; \eref{luOmega} from  $u^\m \eps^{\n\si} Q_{\m\n\si}=0$ .}
\begin{eqnarray}
\fl \dot\phi+\Bigl(\Si-\frac 23\, \theta\Bigr)\Bigl(\ca-\frac12\,\phi\Bigr)-2\,\xi\,\Omega =\cq,\label{luphi}\\
\fl \dot\Si -\frac23\,\dot\theta -\frac12 \,\Bigl(\Si-\frac 23 \,\theta\Bigr)^2 +\ca\,\phi+\ce+2\,\Omega^2= \frac13\,(\m+3\, p -2 \, \Lambda)+\frac12\,\Pi,\label{lusimthe}\\
\fl \dot\ce -\frac 32\,\Bigl( \Si-\frac23\theta\Bigr)\ce -3\ch\xi=\frac 13\dot\m-\frac 12\dot\Pi +\frac 14\Bigl( \Si-\frac23\theta\Bigr)\Pi+\frac 12 \phi\cq-\frac 12(\m+p)\Bigl(\Si-\frac 23 \theta\Bigr),\label{luce}\\
\fl \dot\ch -\frac 32\,\Bigl(\Si-\frac23\theta\Bigr)\ch+3\,\ce\,\xi =\cq\,\Omega +\frac 32\,\Pi\,\xi,\label{luch}\\
\fl \dot \xi -2\,\Bigl(\Si -\frac 16\,\theta\Bigr) \xi  =0 \label{luxi},\\
\fl \dot \Omega -\Bigl(\Si-\frac 23\,\theta \Bigr) \Omega-\ca\,\xi=0 \label{luOmega} .
\end{eqnarray}
\subsubsection*{Transportation}\footnote{Derived as follows. \eref{lusidas} from $u^\m \en^\n u^\si R_{\m\n\si}$}
\begin{eqnarray}
\hat \ca+(\ca+\phi)\ca-\dot\theta-\frac 13\, \theta^2-\frac 32\, \Si^2+2\,\Omega^2= \frac 12 \,(\m+3\,p-2\,\Lambda),\label{lusidas}\\
\dot\m+\theta\,\m +\hat\cq+(2\,\ca+\phi)\cq+ \theta\,p+\frac 32\, \Si\, \Pi =0,\\
\dot\cq+\Bigl(\Si+\frac 43 \, \theta\Bigr)\cq+\hat p +\ca\,p +\hat \Pi+\Bigl(\ca+\frac 32\,\phi\Bigr)\Pi +\m \,\ca =0.\label{rblast}
\end{eqnarray}
Here, the ``dot" derivative is defined $\dot X_{\m\dots\n} := u^\al \na_\al X_{\m\dots\n}$ where $X_{\m\dots\n}$ represents any quantity. The ``hat" derivative is defined $\hat W_{\m\dots\n} := n^\al D_\al W_{\m\dots\n}$, where $W_{\m\dots\n}$ represents a 3-tensor and $D_\m$ is the covariant derivative associated with the 3-sheets.

\section{EM Perturbations on LRS space-times}\label{emPerturbations}

We now consider first-order EM perturbations to arbitrary LRS background space-times defined by \eref{LRSscalars} and \eref{heqs}-\eref{rblast}. The EM perturbations ($E_\m$ and $B_\m$) and the current 3-vector ($J_\m$) are covariant and are considered to be gauge-invariant according to the Sachs-Stewart-Walker lemma \cite{Sachs1964,Stewart1974}.  They are irreducibly split into 1+1+2 form according to
\begin{eqnarray}
\fl E_\m = \emce\, n_\m + \emce_\m ,\qquad B_\m = \emcb \, n_\m +\emcb_\m \qquad\mbox{and} \qquad J_\m = \cj n_\m +\cj_\m.
\end{eqnarray}
The fully non-linear Maxwell equations were previously presented in 1+1+2 form \cite{Clarkson2004}. The corresponding, covariant and gauge-invariant, first-order equations become
\begin{eqnarray}
\fl \hat \emcb+\phi\,\emcb+ \de^\al  \emcb_\al +2\,\Omega \,\emce= 0,\\
\fl \hat \emce+\phi\,\emce+ \de^\al  \emce_\al -2\,\Omega \,\emcb= \emrho \label{BCone},\\
\fl \dot \emcb  -\Bigl(\Si-\frac 23\, \theta\Bigr)\emcb +\epsilon^{\al\be} \de _\al \emce_\be+2\,\emce\,\xi  =0,\\
\fl \dot\emce-\Bigl( \Si-\frac 23\, \theta\Bigr)\emce -\epsilon^{\al\be} \de _\al \emcb_\be-2\,\emcb\,\xi =-\cj,\\
\fl \dot \emcb_{\bar \m}+\Bigl(\frac 12\Si+\frac 23 \theta\Bigr)\emcb_\m- {\epsilon_\m}^\al\Bigl[\hat \emce_\al+\Bigl(\ca+\frac 12\phi\bigr) \emce_\al\Bigr]+ {\epsilon_\m}^\al\de_\al \emce +\Omega {\epsilon_\m}^\al {\emcb}_\al +\xi\emce_\m=0,\label{BCfive}\\
\fl \dot \emce_{\bar \m}+\Bigl(\frac 12\Si+\frac 23 \theta\Bigr)\emce_\m+{\epsilon_\m}^\al\,\Bigl[\hat \emcb_\al+\Bigl(\ca+\frac 12\phi)\emcb_\al\Bigr]- {\epsilon_\m}^\al\,\de_\al \emcb +\Omega {\epsilon_\m}^\al {\emce}_\al -\xi\emcb_\m=- \cj_\m,\nn\\\label{BCsix}
\end{eqnarray}
where in accord with standard notation, a ``bar" over an index implies that the index has been projected onto the 2-sheets. Also, $\emrho$ is the electric charge density and $\epsilon_{\m\n}$ is the anti-symmetric pseudo 2-tensor.

It has long been established that by constructing a complex combination of the EM fields, the system of equations is greatly simplified \cite{Waelsch1913}. This is due to the inherent structure of Maxwell's equations, and this is also true for the fully non-linear equations. They are invariant (in the absence of sources) under the simultaneous transformation $E_\m \rightarrow B_\m$ and $B_\m\rightarrow -E_\m$, which corresponds to $\{\emce \rightarrow \emcb, \emce_\m \rightarrow \emcb_\m \}$ and $\{\emcb \rightarrow -\emce, \emcb_\m \rightarrow -\emce_\m \}$. Therefore, \ref{LADe} uses linear algebra techniques to show that a natural decoupling of the equations is achieved by choosing new dynamical complex quantities according to
\begin{eqnarray}
\Phi     := \emce+\rmi \,\emcb\qquad\mbox{and}\qquad\Phi_\m :=  \emce_\m+\rmi \,\emcb_\m,
\end{eqnarray}
where $\rmi$ is the complex number\footnote{It is also possible to alternatively choose the complex conjugates, $\Phi^*     := \emce-\rmi \,\emcb$ and $\Phi_\m^*     := \emce_\m-\rmi \,\emcb_\m$. Furthermore, any equations governing $\Phi^*$ and $ \Phi_\m^*$ may be found by simply taking the complex conjugate of the equations governing $\Phi$ and $\Phi_\m$.}. Thus, without loss of generality, the six real Maxwell equations \eref{BCone}-\eref{BCsix} are expressed in a new 1+1+2 complex form,
\begin{eqnarray}
\fl \hat \Phi +\phi\,\Phi+ \de^\al\Phi_\al+\rmi\,2\,\Omega\,\Phi=\emrho,\label{divons}\\
\fl \dot\Phi  -\Bigl(\Si-\frac 23\, \theta\Bigr)\Phi+\rmi\,\epsilon^{\al\be}\,\de_\al\Phi_\be+\rmi\,2\,\xi\,\Phi=- \cj,\label{divsec}\\
\fl \dot \Phi_{\bar \m}+\Bigl(\frac 12\Si+\frac 23\theta\Bigr)\Phi_\m-\rmi\,{\epsilon_\m}^\al\Bigl[\hat \Phi_\al+\Bigl(\ca+\frac 12\phi\Bigr)\Phi_\al\Bigr]+\rmi{\epsilon_\m}^\al\de_\al \Phi +\Omega{\eps_\m}^\al \Phi_\al +\rmi\xi\Phi_\m =- \cj_\m. \nn\\\label{thirdone}
\end{eqnarray}

Before proceeding with decoupling the 1+1+2 complex system, we write down the commutation relationships between the various derivatives defined throughout.  These are very important for  the forthcoming analysis and furthermore, it is also vital to perform an integrability check with each and every equation. For any first-order scalar $\Psi$, they were presented previously in \cite{Clarkson2003}
\begin{eqnarray}
\hat{\dot \Psi}-\Bigl(\Si+\frac 13\,\theta\Bigr) \hat \Psi- \dot{\hat \Psi}+ \ca\, \dot \Psi=0, \label{com1}\\
\de_\m \dot \Psi - ( \de_{\bar\m} \Psi)\dot{} +\frac12\,\Bigl(\Si-\frac23\,\theta\Bigr) \de_\m \Psi -\Omega\, {\eps_\m}^\al \de_\al \Psi=0 \label{com2},\\
\de_\m \hat \Psi-  (\de_{\bar\m} \Psi)\hat{} -\frac 12\,\phi\,\de_\m \Psi -\xi\,{\eps_\m}^\al\de_\al \Psi=0, \label{com3}\\
\de_{[\m}\, \de_{\n]}\, \Psi-{\eps_{\m\n}} (\Omega\,\dot\Psi -\xi\,\hat \Psi) =0\label{com4}.
\end{eqnarray}
For a first-order 2-vector, $\Psi_\m$, they were given in \cite{Betschart2004} for LRS class II space-times, and here they are generalized for arbitrary LRS space-times (and they may also be found in the very recent independent study of Clarkson \cite{Clarkson2007}),
\begin{eqnarray}
 \hat{\dot \Psi}_{\bar\m}-\Bigl(\Si+\frac 13\,\theta\Bigr) \hat \Psi_{\bar\m}- \dot{\hat \Psi}_{\bar\m}+ \ca\, \dot \Psi_{\bar\m} -\ch\,{\eps_\m}^\al \Phi_\al=0, \label{com5}\\
 \de_\m \dot \Psi_\n - ( \de_{\bar\m} \Psi_{\bar\n})\dot{} +\frac12\,\Bigl(\Si-\frac23\,\theta\Bigr) \de_\m \Psi_\n -\Omega\, {\eps_\m}^\al \de_\al \Psi_\n=0 \label{com6},\\
 \de_\m \hat \Psi_\n-  (\de_{\bar\m} \Psi_{\bar\n})\hat{} -\frac 12\,\phi\,\de_\m \Psi_\n-\xi\,{\eps_\m}^\al\de_\al \Psi_\n =0, \label{com7}\\
 \de_{[\m}\, \de_{\n]}\, \Psi_\si + K \, \Psi_{[\m} \, \cs_{\n]\si} -\eps_{\m\n} \,\Bigl(\Omega\,\dot\Psi_{\bar\si} -\xi\,\hat \Psi_{\bar\si} \Bigr)=0  \label{com8}.
\end{eqnarray}
Here, the scalar function $K$ has been defined
\begin{eqnarray}
K:=\frac 13\,(\m+\Lambda)-\ce-\frac 12 \,\Pi+\frac 14 \,\phi^2-\frac 14\,\Bigl(\Sigma-\frac 23\,\theta\Bigr)^2 +\xi^2-\Omega^2,\label{gauscurv}
\end{eqnarray}
and this is a natural generalization of the Gaussian curvature scalar defined in \cite{Betschart2004} for LRS class II space-times. In the LRS class II case where $\xi=\Omega=\ch=0$, the sheets mesh to form surfaces for which the Gaussian curvature then has its standard definition \cite{Betschart2004}.

\subsection{Regge-Wheeler equation for LRS space-times}\label{RWLRS}

The, gauge-invariant and covariant, decoupled equation governing $\Phi$ is derived by taking the ``dot" derivative of \eref{divsec} and it is important to use  the Ricci/Bianchi identities \eref{rb1}-\eref{rblast}, the scalar function $K$  \eref{gauscurv}, and the commutation relationships for the various derivatives \eref{com1}-\eref{com8}. It is also necessary to substitute \eref{divons}-\eref{thirdone}  for further simplifications and after some arduous manipulation, we arrive at
\begin{eqnarray}
\ddot \Phi   -\Bigl(\Si-\frac 53\, \theta -\rmi\,2\,\xi\Bigr)\dot \Phi -\hat{\hat\Phi} -(\ca+2\,\phi+\rmi\,2\,\Omega)\hat \Phi - V\Phi =\mc{S}.\label{GRW}
\end{eqnarray}
The potential and energy-momentum source have been defined
\begin{eqnarray}
\fl V :=\de^2+ 2\,K-\m+p+\Pi -2\,\Lambda +\rmi\,4 \,\Bigl[\Omega\,\ca -\xi \,\Bigl(\Si+\frac13\,\theta\Bigr) \Bigr]\label{potentialLRS} ,\\
\fl \mc{S} := - \hat \emrho -(\phi+\ca) \emrho-\dot \cj  -\theta\,\cj +\rmi\,\epsilon^{\al\be} \de_\al \cj_\be,
\end{eqnarray}
 where the 2-Laplacian is $\de^2 := \de^\al \de_\al$.  This is a new complex RW equation generalized for EM perturbations on arbitrary LRS space-times, and this generalizes the RW equation derived in \cite{Betschart2004} for LRS class II space-times.
 
 By inspecting \eref{GRW}, this clearly demonstrates that for arbitrary LRS space-times, the complex EM scalar, $\Phi$, decouples from the complex EM 2-vector, $\Phi_\m$.  It also indicates that the radial {\it electric} field ($\emce$) and radial {\it magnetic} field ($\emcb$) do not decouple from each other and instead they must be treated as a single complex radial {\it electromagnetic} field ($\Phi$).  We show in the next section that further decoupling can be achieved in specific sub-cases.

\subsubsection{LRS class II space-times}\label{LRSII}

A closer inspection of the coefficients in the complex RW equation \eref{GRW} reveals that the imaginary components are always associated with either $\xi$ or $\Omega$. Therefore, further decoupling is achieved in the case when they vanish and by \eref{heqs} this implies $\ch$ will also vanish, and this is precisely LRS class II defined by
\begin{eqnarray}
\mbox{LRS class II:}\,\, \{\ca, \,\theta,\, \phi,\,\Si,\,\ce ,\,\m,\, p,\,\cq,\,\Pi,\, \Lambda\}.
\end{eqnarray}
Thus, for LRS class II the complex RW equation \eref{GRW} reduces to
\begin{eqnarray}
\ddot \Phi   -\Bigl(\Si-\frac 53\, \theta \Bigr)\dot \Phi -\hat{\hat\Phi} -(\ca+2\,\phi)\hat \Phi - V\Phi =\mc{S},\label{GRWLRSII}
\end{eqnarray}
where the potential is now,
\begin{eqnarray}
V :=\de^2+ 2\,K-\m+p+\Pi -2\,\Lambda \label{potential} ,
\end{eqnarray}
which is a remarkably simple form; for example, in vacuum space-times whereby the cosmological constant vanishes, the potential is {\it purely} in terms of the Gaussian curvature of the 2-sheets (and the 2-Laplacian).

Now since all differential operators (along with their coefficients) acting on $\Phi$ in \eref{GRWLRSII}  are purely real, there are two independent decoupled equations here; one for each of the real and imaginary components of $\Phi$ (i.e. $\emce$ and $\emcb$),
\begin{eqnarray}
\fl \ddot \emce   -\Bigl(\Si-\frac 53\, \theta\Bigr)\dot \emce -\hat{\hat\emce} -(\ca+2\,\phi)\hat \emce - V\,\emce =- \hat \emrho -(\phi+\ca) \emrho-\dot \cj  -\theta\,\cj \label{waveforce},\\
\fl  \ddot \emcb   -\Bigl(\Si-\frac 53\, \theta\Bigr)\dot \emcb -\hat{\hat\emcb} -(\ca+2\,\phi)\hat \emcb - V\,\emcb=\epsilon^{\al\be} \de_\al \cj_\be
\end{eqnarray}
and these correspond to those derived in \cite{Betschart2004}. However, these equations correct an error that resides in the potential of the work presented by \cite{Betschart2004}, for which we now elucidate. For better comparison with \cite{Betschart2004}, we can substitute \eref{gauscurv} into the potential \eref{potential} to reveal,
\begin{eqnarray}
V =\de^2+\frac 12 \,\phi^2-2\,\ce+\Bigl(\frac29\,\theta-\frac 13\, \Sigma\Bigr)\Bigl(\frac 32\, \Sigma-\theta\Bigr)-\frac13\,(\m-3\,p+4\,\Lambda) .\label{potential1}
\end{eqnarray}
We will denote their incorrect potential as $V_{BC}$ and reproduce this from \cite{Betschart2004},
\begin{eqnarray}
V_{BC} =\de^2 +\frac 12 \,\phi^2-2\,\ce\underbrace{+\Bigl(\frac 13\, \theta+ \Sigma\Bigr)\Bigl(\frac 32\,\Si-\theta\Bigr)}_{\mbox{incorrect term}}-\frac13\,(\m-3\,p+4\,\Lambda).
\end{eqnarray}
However, it is strongly emphasized here that this in no way affects the way in which the equations decouple in  \cite{Betschart2004}. Furthermore, \cite{Betschart2004} presents an informative and interesting analysis of various applications for which the ``incorrect term" vanishes, and thus those results remain intact.

\subsection{Bardeen-Press equations for LRS space-times}\label{BardeenPressLRS}

We now show that since we are exploiting the inherent structure of the equations,  we can derive a new decoupled equation for  the complex EM 2-vector, $\Phi_\m$. The derivation is similar to  how the generalized RW equation was constructed. First take the dot derivative  of \eref{thirdone} and use \eref{rb1}-\eref{com8}, and substitute \eref{divons}-\eref{thirdone}, to simplify further. Ultimately, we find a  decoupled, covariant and gauge-invariant,  equation given by
\begin{eqnarray}
\ddot \Phi_{\bar \m} - \Bigl(\Si-\frac 53 \,\theta-\rmi\,2\,\xi\Bigr)\dot \Phi_{\bar\m}-\hat{\hat \Phi}_{\bar\m} -(\ca+2\,\phi+\rmi\,2\,\Omega)\, \hat \Phi_{\bar\m}-V_{{{(1)}}} \Phi_\m \nonumber\\
  -\rmi \,{\epsilon_\m}^\al \,\left[(2\,\ca-\phi +\rmi\,2\,\Omega)\dot \Phi_\al   -(3 \, \Si +\rmi\,2\,\xi)\,\hat \Phi_\al -V_{(2)}\, \Phi_\al\right]=\mc{S}_\m,\label{waveforphia}
\end{eqnarray}
where two terms related to the potentials have been defined
\begin{eqnarray}
\fl V_{(1)}:=\de^2+\ce+\frac14\phi^2-\ca^2+\phi\ca +\frac 74\, \Si^2-\frac 29\,\theta^2+\frac 23\,\theta\,\Si-\frac 13\,\m+p-\frac 43\Lambda +\xi^2-\Omega^2 \nn\\
+\rmi \Bigl[\Omega\,\Bigl(2\,\ca+\phi\Bigr) -\xi \Bigl(\Si+\frac 43\,\theta\Bigr) -\ch\Bigr],\\
\fl V_{(2)} := -\dot\ca+\hat\theta+\frac 23\,\theta\,(\phi-2\,\ca) +\frac 12\,\Sigma\,(\phi+4\,\ca)-2\,\xi\,\Omega -\cq \nn\\
+\rmi\,\Bigl[ 2\,\Omega\,\Bigl(\Si+\frac 13\,\theta\Bigr) -2\,\xi\,\ca \Bigr], 
\end{eqnarray}
and the energy-momentum source
\begin{eqnarray}
\mc{S}_\m:=-\dot \cj_{\bar\m}+\frac32\,\Bigl(\Si-\frac 23 \, \theta\Bigr) \cj_\m-\de_\m \emrho+\Omega\,{\eps_\m}^\al \cj_\al  \nn\\
+\rmi\, {\epsilon_\m}^\al \Bigl(\de_\al \cj -\hat \cj_\al-\frac32\,\phi\, \cj_\al \Bigr) -\rmi\, \xi\,\cj_\m.
\end{eqnarray}
It is now clear that the complex EM 2-tensor also decouples from the complex EM scalar $\Phi$.  Thus analogous to the radial case, the {\it electric} 2-vector ($\emce_\m$) and the {\it magnetic} 2-vector ($\emcb_\m$) do not decouple from each other, however,  they combine to form a single complex {\it electromagnetic} 2-vector ($\Phi_\m$).

In the next section we show how to further decompose \eref{waveforphia} to find two new BP equations generalized for EM perturbations to LRS space-times.

\subsubsection{$1+1+1+1$ Decomposition}\label{1111}

In order to decouple the two components residing in \eref{waveforphia}, we consider a further projection along two more vectors. The natural decoupling methodology in \ref{LADe} is employed again, and this allows these vectors to be a complex-conjugate pair $(m^\m, m^{*\m})$ which satisfies the following relationships:
\begin{eqnarray}
 m^{*\al} m_\al =1,\qquad m^\al m_\al =0, \qquad { m^{*\al} }{m^*_\al} =0,\qquad \cs ^{\m\n} = 2\, m^{(\m} {\bar m}^{*\n)},\label{mbarprops}
\end{eqnarray}
where $\cs_{\m\n}$ is the projection tensor for the 2-sheets and these complex-conjugate vectors are orthogonal to both $u^\m$ and $n^\m$.  Consider the arbitrary 2-vector, $\Psi_\m$, and 2-tensor, $\Psi_{\m\n}$,  then by subsequently using \eref{mbarprops}, a new irreducible decomposition into 1+1+1+1 form is given by,
\begin{eqnarray}
\fl \Psi_\m = (\Psi_\al m^\al) m^*_\m + ( \Psi_\al  m^{*\al}) m_\m,\label{deompofum}\\
\fl  \Psi_{\m\n} = ( \Psi_{\al\be}  m^{*\al}  m^{*\be})  m_\m  m_\n+( \Psi_{\al\be} m^\al m^\be)  m^*_\m m_\n^*+ 2\, m_{(\m}  m^*_{\n)}\,(\Psi_{\al\be}  m^{*(\al} m^{\be)})\nonumber\\
+2\, m_{[\m} \bar m_{\n]}\,(\Psi_{\al\be}  m^{*[\al} m^{\be]}) ,\label{antisyshedec}
\end{eqnarray}
where the individual components are scalars with a specific spin-weight \cite{Penrose1987} that has a standard definition as follows. Let the complex vector, $m^\m$, undergo a transformation on the 2-sheet according to $
m^\m \rightarrow \mc{C} \,{\mc{ C}}^{*-1}\, m^\m$ where $\mc{C}$ is an arbitrary complex scalar field and $ \mc{C}^*$ its complex conjugate. Then any quantity, ${\zeta_{\m\dots\n}}^{\si\dots\ta}$,  which has a corresponding transformation of
\begin{eqnarray}
{\zeta_{\m\dots\n}}^{\si\dots\ta}\rightarrow \mc{C}^p\,  \mc{C} ^{*q}\,{\zeta_{\m\dots\n}}^{\si\dots\ta},
\end{eqnarray}
is said to have a spin-weight $s$ defined,
\begin{equation}
s:= \frac 12\,(p-q).
\end{equation}
We now derive the new and important quantities which arise from the further decomposition of the 1+1+2 formalism into the 1+1+1+1 formalism. All the subsequent equations will naturally occur in complex-conjugate pairs. However, we only display one of the pair and note that the other is found by taking the complex conjugate. Thus we have
\begin{eqnarray}
\de_\m m_\n = -m_\m m_\n  m^{*\al} \chi_\al -\frac 12\,\cs_{\m\n} m^\al \chi_\al -\rmi\,\frac 12\,\si \,m^\al \chi_\al \,\epsilon_{\m\n} \label{sheedecopdodm},
\end{eqnarray}
where
\begin{eqnarray}
 \chi_\m := m^\al \de_\m m^*_\al=- \chi^*_\m
\end{eqnarray}
 is purely imaginary and has zero spin-weight, and
\begin{eqnarray}
\si:=\rmi\, m_\al  m^*_\be \epsilon^{\al\be} \qquad\mbox{from which it follows}\qquad\si^2=1,
\end{eqnarray}
therefore, $\si$ has zero spin-weight and is purely real. Thus, \eref{sheedecopdodm} depends only on the complex-conjugate pair $(m^\al \chi_\al,m^{*\al} \chi^*_\al)$ which have a spin weight of $1$ and $-1$ respectively. We also have a constraint and a relationship for the divergence, which are respectively
\begin{eqnarray}
(\de^\al+\chi^\al )m_\al =0\qquad\mbox{and}\qquad\epsilon^{\al\be} \de_\al m_\be= \rmi\,\si\, \de^\al m_\al,
\end{eqnarray}
and finally, by using \eref{antisyshedec}, the Levi-Civita psuedo 2-tensor is decomposed as
\begin{eqnarray}
\epsilon_{\m\n}=\rmi\,2\,\si\,m_{[\m}  m^*_{\n]}.
\end{eqnarray}

\subsubsection{Bardeen-Press equations for scalars spin weighted scalars}
We now have developed the necessary mathematical tools to irreducibly  decomposed the complex 2-vector as
\begin{eqnarray}
\Phi_\m  =  \mop\, { m^*_\m} + \moc\,m_\m,\label{angdeophiq}
\end{eqnarray}
where $\mop:= m^\al \Phi_\al$ has a spin-weight of $s=1$ and $\moc:=  m^{*\al} \Phi_\al$ has a spin-weight of $s=-1$. Finally, by substituting \eref{angdeophiq} into \eref{waveforphia} and contracting separately with $m^\m$ and $ m^{*\m}$, we find the components naturally decouple into two spin-weighted equations of the form,
\begin{eqnarray}
\fl\ddot\mop+\Bigl[2\,\ga -\Si+\frac53\,\theta+\rmi\,2\,\xi-s\,\si\,(2\,\ca-\phi-\rmi\,2\,\Omega)\Bigr]\dot \mop - \hat{\hat \mop}  \nn\\
\fl-[2\,\lambda+\ca+2\,\phi-s\,\si\,(3\,\Si-\rmi\,2\,\xi)]\hat\mop -2\, \chi^\al \de_\al \mop -V_{B}\,\mop=S_\oplus\label{BPones},
\end{eqnarray}
\begin{eqnarray}
\fl\ddot \moc+\Bigl[2\,\bar \ga -\Si+\frac53\,\theta+\rmi\,2\,\xi-s\,\si\,(2\,\ca-\phi-\rmi\,2\,\Omega)\Bigr]\dot \moc - \hat{\hat {\moc}} \nn\\
\fl -[2\,\bar\lambda+\ca+2\,\phi-s\,\si\,(3\,\Si-\rmi\,2\,\xi)]\hat \moc-2\, \bar\chi^\al \de_\al \moc- V_{P}\moc= S_{\otimes}\label{BPtwos}.
\end{eqnarray}
where the energy-momentum source has similarly been decomposed as \mbox{$S_\m:= S_\oplus \, m^*_\m+ S_\otimes \,m_\m$}. Furthermore, some new definitions involving various combinations of $m^\m$ and $m^{*\m}$ are, 
\begin{eqnarray}
\fl \ga := m^\al \dot {m}^*_\al,\qquad \lambda := m^\al \hat{ \bar{m}}^*_\al,\qquad \chi_\m := m^\al \de_\m m^*_\al\qquad\mbox{and}\qquad \chi := (\de^\al m^\be)(\de_\al m^*_\be),
\end{eqnarray}
and the terms related to the potentials are now
\begin{eqnarray}
 \fl V_{B}:= -\dot\ga -\ga\Bigl[\ga -\Si+\frac 53\,\theta -s\,\si\,(2\,\ca-\phi)\Bigr]+\hat \lambda +\lambda\,(\lambda+\ca+2\,\phi-3\,s\,\si\,\Si) \nonumber\\
\qquad -\chi+\de^\al\chi_\al+V_{(1)}-\si\,V_{(2)},\\
\fl V_{P}:=-\dot{ \ga}^* - \ga^*\Bigl[ \ga^* -\Si+\frac 53\,\theta -s\,\si\,(2\,\ca-\phi)\Bigr]+\hat { \lambda }^*+ \lambda^*\,( \lambda^*+\ca+2\,\phi-3\,s\,\si\,\Si) \nonumber\\
\qquad-\bar \chi+\de^\al \chi^*_\al+V_{(1)}+\si\,V_{(2)}.
\end{eqnarray}
The decoupled equations, \eref{BPones}-\eref{BPtwos}, are new generalizations of the BP equations for arbitrary LRS space-times. The generalized BP equations \eref{BPones}-\eref{BPones} were checked that they reduce to the original Schwarzschild result, which was derived using the Newman-Penrose formalism, as presented in \cite{Chandra1983}. The equations were expressed in coordinate form using Maple 9.5 and they correspond precisely.

\section{Summary and Conclusions}

We have successfully decoupled, gauge-invariant and covariant, EM perturbations on arbitrary LRS space-times.  We used an eigenvector/eigenvalue analysis to take advantage of the inherent mathematical characteristics of Maxwell's equations and express them in a complex 1+1+2 form that facilitates decoupling. This new complex system  was then used to demonstrate the decoupling of the complex EM scalar ($\Phi$) and the complex EM 2-vector $(\Phi_\m)$.  The governing equation for $\Phi$ is a RW equation generalized towards arbitrary LRS space-times. Furthermore, we also derived a new decoupled equation governing the complex EM 2-vector.  We then developed a further decomposition of the 1+1+2 formalism into a 1+1+1+1 formalism and ultimately derived a pair of decoupled spin-weighted scalars. The governing equations are the BP equations generalized for arbitrary LRS space-times. Finally, we also noted that additional decoupling could be achieved between the EM scalars, $\emce$ and $\emcb$, by reducing the RW equation to the LRS class II sub-case.

This process presented here is also highly useful as a mathematical guide for decoupling the analogous case of gravitational perturbations to LRS space-times using the 1+1+2 {\it gravito-electromagnetic} (GEM) formalism. We have already shown how to decouple complex GEM spin-weighted scalars for the covariant Schwarzschild case \cite{Burston2006} and we will show in a future paper that this can be extended for general LRS space-times.
\newline

\appendix

\section{Linear Algebra: Decoupling Systems of Differential Equations}\label{LADe}

Consider the system given by, 
\begin{eqnarray}
L_1 \,E+L_2\,B =0\mbox{    and     } L_1\,B-L_2\,E=0,
\end{eqnarray}
where $L_1$ and $L_2$ represent differential operators and $E$ and $B$ are any scalar fields.  This system has the property that it is invariant under the simultaneous transformation of $E\rightarrow B$ and $B\rightarrow-\,E$. This system can be expressed in  a matrix form as
\begin{eqnarray}\label{matrix_one}
 \left(\begin{array}{c}
L_1\,E     \\
L_1\,B 
\end{array}\right) + M \left(\begin{array}{c}
L_2\,E     \\
L_2\,B 
\end{array}\right)= \left(\begin{array}{c}
 0         \\
    0 
\end{array}\right), \mbox{  where     } M:=\left(\begin{array}{cc}
 0     &1    \\
   -1& 0 
\end{array}\right)
\end{eqnarray}
is the matrix responsible for coupling $E$ to $B$. $M$ can be written in terms of its eigenvalues, $\mbox{diag}({D})$, and corresponding eigenvectors, col($P$), according to $M = P \, D\,P^{-1}$.
Therefore, since $D$ is diagonal, it is clear that by multiplying  (\ref{matrix_one}) by $-2\,\rmi\,P^{-1}$ results in the decoupled system,
\begin{eqnarray}
\fl  L_1(E+\rmi\,B)+\rmi\,\,L_2(E+\rmi\,B)=0\qquad\mbox{and}\qquad L_1(E-\rmi\,B)-\rmi\,\,L_2(E-\rmi\,B)=0.
\end{eqnarray}
Thus a complex-conjugate pair of equations arise. This result can be generalised to tensors of any type without loss of generality provided the invariance is satisfied. The matrix $M$ needs to be written in block form with blocks of zeros or the identity matrix to compenstate for the number of dimensions.

\end{document}